\magnification=\magstep1
\font\bigbfont=cmbx10 scaled\magstep1
\font\bigifont=cmti10 scaled\magstep1
\font\bigrfont=cmr10 scaled\magstep1
\vsize = 23.5 truecm
\hsize = 15.5 truecm
\hoffset = .2truein
\baselineskip = 14 truept
\overfullrule = 0pt
\parskip = 3 truept
\def\frac#1#2{{#1\over#2}}

\input epsf
\nopagenumbers

\topinsert
\vskip 3.2 truecm
\endinsert
\centerline{\bigbfont FERMIONS UNDER STRONG REPULSION:}
\vskip 6 truept
\centerline{\bigbfont FROM MULTICONNECTED FERMI SURFACES}
\vskip 6 truept
\centerline{\bigbfont TO FERMI CONDENSATION}

\vskip 20 truept

\centerline{\bigifont Yu. G. Pogorelov and V. R. Shaginyan}
\vskip 8 truept
\centerline{\bigrfont Centro de F\'{\i}sica do Porto, Universidade do Porto,}
\centerline{\bigrfont 4169-007 Porto, Portugal}
\vskip 2 truept
\centerline{\bigrfont and }
\vskip 2 truept
\centerline{\bigrfont Petersburg Nuclear Physics Institute, RAS,}
\centerline{\bigrfont 188300 Gatchina, Russia}

\vskip 1.8 truecm

\centerline{\bf 1.  INTRODUCTION}
\vskip 12 truept

The studies of strongly interacting fermion systems are of
considerable interest in solid state physics, nuclear physics, 
astrophysics, and so on. It is well known that many of the basic 
properties result already from the notion of the quasiparticles, 
at temperature $T=0$, characterized by the step-like occupation 
function $n({\bf p},\sigma) = n_F({\bf p},\sigma) = \theta (p_F - 
p)$, so that the number density,  $\rho = \sum_{\sigma} \int n({
\bf p},\sigma) d{\bf p}/(2 \pi)^3$, determines the Fermi momentum
$p_F^3 = 3\pi^2\rho$ [1].

In real physical systems, presence of various interactions between
particles (by Coulomb forces, mesonic forces, van der Vaals forces, 
etc.) can essentially change the ground state energy and can rarely 
diversify the phenomenology of a Fermi system, preserving it as a 
Landau Fermi liquid. The Landau theory of normal Fermi liquids is 
based on the notions of the quasiparticles and the amplitudes $F$ 
which characterize the effective interaction between the quasiparticles 
[1]. It permits to describe all the observable physical properties in 
terms of the first and second variational derivatives of the energy 
functional $E[n(\bf {p},\sigma)]$ with respect to the occupation numbers 
$n(\bf {p}, \sigma)$: $\varepsilon({\bf p},\sigma) = \delta E/\delta 
n(\bf {p},\sigma)$ and  $F({\bf p}, \sigma,{\bf p}^\prime, \sigma^ 
\prime ) = \delta^2 E/\delta n({\bf p}, \sigma) \delta n({\bf p}^ 
\prime, \sigma^\prime)$. As a result, the Landau theory has removed 
high energy degrees of freedom and kept a sufficiently large number of 
relevant low energy degrees of freedom to treat liquid's low energy 
properties. Usually, it is assumed that the breakdown of the Landau 
theory is defined by the Pomeranchuk stability conditions and occurs 
when the Landau amplitudes being negative reach its critical value [2]. 
The new phase at which the stability conditions are restored can in 
principle be again described in the framework of the Landau theory. 
But it is not a general case. The most evident example is the 
superconducting transition under arbitrarily weak attraction between 
fermions with opposite spins $\sigma$ and opposite momenta close to the 
Fermi momentum $p_{\rm F}$ [3]. But one can also search for alternative
scenarios of such restructuring produced by repulsive interactions,
if strong enough. The quasiparticle dispersion law
$$
\varepsilon({\bf p},\sigma)={{\delta E}\over
{\delta n({\bf p},\sigma)}}\eqno(1)
$$
satisfies the stability criterion: $v_F = [\partial \varepsilon
({\bf p},\sigma)/\partial p]_{p = p_F} > 0$. Then the minimum of the
energy,
$$
\delta \left(E - \mu \sum_{\sigma} \int n({\bf p},
\sigma) {{d{\bf p}}\over{(2 \pi)^3}}\right)=
\int \left(\varepsilon({\bf p},\sigma)-\mu\right)
\delta n({\bf p},\sigma)
{{d{\bf p}}\over{(2 \pi)^3}}=0, \eqno(2)
$$
is reached (although with the chemical potential $\mu$ generally
different from $p_F^2/2m$) still for $n = n_F$.
This quasiparticle occupation should not be confused with the
renormalization of {\it real particle} occupation function [4].

However one can also search for more general solutions of Eq. (2),
$n \neq  n_F$. Regarding the Pauli principle restriction
$0 \leq n({\bf p}, \sigma) \leq 1$, these general solutions must
satisfy the equation
$$
n({\bf p},\sigma) [1 - n({\bf p},\sigma)]
[\varepsilon({\bf p},\sigma) - \mu] = 0.  \eqno(3)
$$
The possibility for the last factor in the l.h.s. of Eq. (3) to be
zero implies a possibility for the so-called Fermi condensate (FC),
manifested by a flat dispersion law: $\varepsilon({\bf p},\sigma)
\equiv \mu$, and a fractional occupation $0 < n({\bf p}, \sigma)
< 1$ within a certain finite spherical layer $p_i \leq p \leq p_f$
in the momentum space. Otherwise, the occupation function can only
take the values 0 or 1, so that  $\varepsilon({\bf p},\sigma) <
\mu$, $n({\bf p}, \sigma) =1$ at $p < p_i$ and $\varepsilon({\bf
p},\sigma) > \mu$, $n({\bf p}, \sigma) = 0$ at $p > p_f$ [3,5-7].
But the full set of alternatives for the ground state includes,
besides FC and Fermi ground state (FGS), also the so-called
multiconnected Fermi states (MFS). This possibility was first
recognized yet in the Hartree-Fock framework [8,9]. We will study 
such a possibility using model forms for the interaction:  $F({\bf p}, 
\sigma,{\bf p}^\prime, \sigma^\prime) = g_{\sigma,\sigma^\prime} 
U_{\sigma, \sigma^\prime}(q)$, with only relevant dependence on the 
momentum transfer $q=|{\bf p-p}^\prime|$ and some coupling constants 
$g_{\sigma,\sigma^\prime}$.

Let us restrict consideration to uniform, isotropic, and spin-
symmetric Fermi systems, where $\varepsilon({\bf p},\sigma)=
\varepsilon(p)$ (this is also the scope in which all the known FC
models were considered). Then any MFS presents a sequence of fully
occupied areas (we call them icebergs) where $\varepsilon(p) < \mu$,
separated by empty spacers where $\varepsilon(p) > \mu$. Hence it
can be uniquely labeled by the number $N$ of interfaces between
icebergs and spacers: MFS$(N)$. When this number is odd, the
MFS$(N=2n-1)$ consists of $n$ icebergs: $p_{2i-1} < p < p_{2i}$
($ i = 1\, ,\dots\, n$), and $n$ spacers: $ p_{2i-2} <  p <
p_{2i-1}$ (including a central spherical void $0 < p < p_1$). For
even numbers, the MFS$(N=2n)$ contains $n$ icebergs (including a
central spherical island $0 < p < p_1$) and $n-1$ spacers. Interfaces
$p_i$ are fixed by the equilibrium conditions
$$
\varepsilon(p_i)= \mu\, ,\quad i = 1\, ,\dots\, N\, , \eqno(4a)
$$
and icebergs fit to the normalization condition:
$$
\sum_{i=1}^n (p_{2i}^3 -  p_{2i-1}^3) =  3 \pi^2 \rho  \eqno(4b)
$$
(for odd $N$, the analogue for even $N$ is evident). In this
context, FGS can be identified as MFS$(0)$.

At least, the choice between the two different types of FGS
restructuring with growing repulsion constant $g$ is uniquely
defined by the analytic properties of the model potential $U(q)$,
so that FGS$\to$FC scenario is only possible if this function
(or its derivatives) have singular points at real axis in the
complex $q$-plane, otherwise the infinite sequence of transitions
FGS$\to$MFS$(1) \to$MFS$(2) \to \dots$  takes place [10].

Despite such strict dichotomy between the two evolution routes of
model fermion systems, several analytic and numerical studies [10,11]
showed that their expected physical properties should become very
close in this course. If the distinction between analytic and
non-analytic potentials is not fully respected, one can even arrive
at a wrong conclusion on possible MFS$(N) \to$FC transition at some
finite $N$ [12]. In fact, as will be shown below, the FC state is
only reached in the limit of MFS$(N \to \infty)$ (for a given
analytic $U(q)$). But aside of these formal issues, there remains
a clear interest in practical conditions when the difference between
the MFS and FC states cannot be ignored, permitting to use
the MFS picture for explanation of various actual physical effects
related to the behavior of heavy-fermion systems in magnetic
fields [13,14]. The present work is just focused on a detailed
analysis of common features between FC and advanced MFS$(N\gg1)$
states.  To this end, we use the analytic model $U(q) = 1/\sqrt
{q^2+q_0^2}$ which tends to the non-analytic FC model $1/q$ in the
limit $q_0 \to 0$, to develop an effective approximation for the
average occupation function in this limit and to establish the
criteria on model parameters ($g$, $q_0$) and external parameters (as
temperature) for the two states to be practically indistinguishable.
\vskip 28 truept

\centerline{\bf 2.  ANALYTIC SOLUTION FOR THE $1/\sqrt {q^2+q_0^2}$
\bf MODEL}
\vskip 12 truept

Let us start from the model interaction
$$
F({\bf p},\sigma,{\bf p}^\prime,\sigma^\prime)={g\over{\sqrt{(
{\bf p-p}^\prime)^2 + q_0^2}}}.  \eqno(5)
$$
This form implies spherically symmetric and spin independent
occupation: $n({\bf p}, \sigma) \to n(p)$, and we can present the
energy functional as a 1D integral:
$$
E[n(p)] = {1\over \pi^2}\int p^2  n(p)\left[\varepsilon_0(p)+V(p)/2\right]
{\rm d}p.  \eqno(6)
$$
The potential energy $V(p)$ also enters the quasiparticle dispersion law:
$$
\varepsilon({\bf p},\sigma) = \varepsilon_0(p) + V(p)\, , \eqno(7)
$$
here $\varepsilon_0(p)=p^2/(2m)$ is the kinetic energy of system's
particle with the bare mass $m$.
$V(p)$ presents a function of $p$ and a functional of $n(p^\prime)$ given
by a 1D integral:
$$
V(p)= {g\over \pi^2} \int {p^\prime}^2 n(p^\prime)v(p,p^\prime)
{\rm d}p^\prime,  \eqno(8)
$$
where the kernel $v(p,p^\prime)$ results from integration of the model
potential $U(|{\bf p - p}^\prime|)$ in the solid angle between the
vectors ${\bf p}$ and ${\bf p}^\prime$:
$$ v(p,p^\prime) = {1\over{2
p p^\prime}}\int_{|p-p^\prime|}^{p+p^\prime} t U(t) {\rm d}t.
\eqno(9) $$

In the limit $q_0 \to 0$, the model, Eq. (5), passes into the FC model
$U_{\rm FC}(q) = 1/q$ [5,7]. Recall briefly its main properties in the
present notation. The kernel, Eq. (9), is here simply
$$
v_{\rm FC}(p,p^\prime) = 1/\max (p,p^\prime)\, , \eqno(10)
$$
At small enough $g$, when FGS still holds, this gives the explicit
potential energy
$$
V_{\rm FC}(p) = \cases { 3 \xi p_{\rm F}^2/(2 m) - \xi p^2/(2 m)\, ,
&$\quad p < p_{\rm F}\,$, \cr
& \quad \cr
\xi p_{\rm F}^3/(m p)\, ,\quad & $ \quad p > p_{\rm F}\,$. \cr } \eqno(11)
$$
When $\xi=g m/(3 \pi^2) \to 1$, the dispersion in Eq. (7) vanishes for all
occupied states, manifesting the onset of FC, which leads to the
chemical potential $\mu_{\rm FC}$ that is lower then the
corresponding to FGS.  For $\xi > 1$, the FC occupation function is:
$n_{\rm FC}(p) = \xi^{-1} \theta(p_f - p)$ with $p_f = p_{\rm F}
\xi^{1/3}$, thus keeping the quasiparticle energy constant:
$\varepsilon(p)=3 p_f^2/(2m)$, within the whole occupied band $p <
p_f$ ($p_i=0$ in the $1/q$ model).

In the actual model, Eq. (5), the kernel takes a more complicate form
$$
v(p,p^\prime)={{\sqrt{(p+p^\prime)^2+q_0^2}-\sqrt{(p-p^\prime)^2+
q_0^2}} \over{2 p p^\prime}}\, ,
$$
and using it in Eq. (8) at small enough $g$ when FGS holds, we
calculate the potential energy produced by the fully occupied Fermi
sphere of radius $p_{\rm F}$:
$$
V(p) = V_{p_{\rm F}}(p)+V_{p_{\rm F}}(-p)\, , \eqno(12)
$$

where
$$
V_P(p)=\frac{\xi}{2 m}\left[\frac{\sqrt{(P+p)^2+q_0^2}}{p}\left(P^2 +
q_0^2+p\frac{P-p}{2}\right)-\frac{3q_0^2}{2}\,{\rm arcsinh}{{P+p}
\over q_0}\right]\, .
$$
\epsfbox{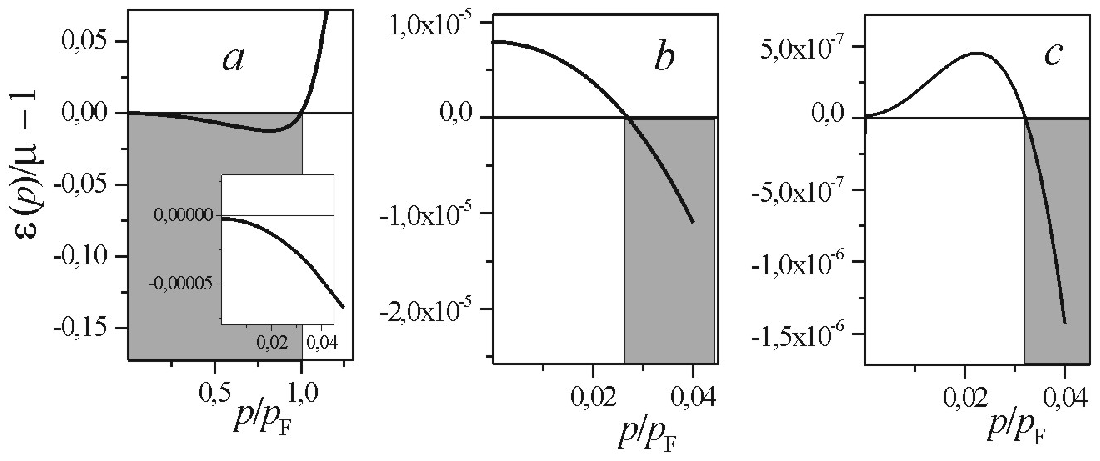}

{\bf Figure 1.} Instabilities of Fermi states in the
$\xi/\sqrt{q^2+q_0^2}$ model at $q_0 = 0.1 p_{\rm F}$ with growing
$\xi$. a) Dispersion law $\varepsilon(0)$ tends to $\mu$ from below
at $\xi \to \xi_1 \approx 1.04293$. b) A void is growing in the center
of filled Fermi sphere at $\xi_1 < \xi <\xi_2$, c) an island will appear
in the center of the void at $\xi = \xi_2 \approx 1.04356$.
\vskip 12 truept
Note that, unlike Eq. (11), the form of Eq. (12) defines {\it analytic}
behavior of $\varepsilon(p)$ for all real $p$. In presence of finite
``screening parameter'' $q_0$, the repulsion potential $U(q)$ is reduced
compared to $U_{\rm FC}(q)$, hence the critical value of $\xi$, when
FGS fails, turns higher than 1. It is easy to see that $V(p)$ is an even 
function of $p$ with maximum at $p=0$ and satisfies the inequality: $V(p) 
> V(0) + p^2 V^{\prime \prime}(0)/2$. Hence the instability of FGS in 
this model occurs just in the center of the Fermi sphere, when 
$\varepsilon(0) \to\varepsilon (p_F) = \mu$ (Fig. 1a). As discussed 
below, this instability is the first in an infinite series, and the 
above condition involves the corresponding values of potential energy:
$$
V(0)=  \frac{3\xi}{2 m}\left(p_{\rm F}\sqrt{p_{\rm F}^2+q_0^2}-q_0^2
{\rm arcsinh}\frac{p_{\rm F}}{q_0}\right),
$$
and
$$
V(p_F)=\frac{\xi}{2m}\left[(p_{\rm F}^2+q_0^2)\frac{\sqrt{4p_{\rm F}^2+
q_0^2}}{p_{\rm F}}-\frac{3q_0^2}{2}{\rm arcsinh} \frac{2 p_{\rm F}}{q_0}-
\left(\frac{q_0}{p_{\rm F}}\right)^3\right]\, .
$$
Using them in Eq. (1) we obtain the first critical value of $\xi$ as a
function of $q_0$:
$$
\eqalignno{
\xi_1(q_0) = & [3\frac{\sqrt{p_{\rm F}^2+q_0^2}}
{p_{\rm F}}-\frac{( p_{\rm F}^2+q_0^2 )
\sqrt{4p_{\rm F}^2+q_0^2}}{p_{\rm F}^3}   \cr
&   +\frac{3 q_0^2}{2p_{\rm F}^2} \left({\rm arcsinh}
\frac{2 p_{\rm F}}{q_0}-2{\rm arcsinh}\frac{p_{\rm F}}{q_0} \right)+
\left(\frac{q_0}{p_{\rm F}}\right)^3 ]^{-1}\,.  &(13) \cr }
$$
It behaves as
$$
\xi_1(q_0) \approx 1+ \left(\frac{3q_0}{2p_{\rm F}}\right)^2
\ln \left(\frac{p_{\rm F}}{q_0} \right)  \eqno(14)
$$
in the important limit $q_0 \ll p_{\rm F}$ and as $\xi(q_0)
\approx (q_0/p_{\rm F})^3 $ at $q_0 \gg p_{\rm F}$. At $\xi \to
\xi_1(q_0)$, the dispersion law in the center is characterized by
a {\it negative} effective mass: $m^* = 1/\varepsilon^{\prime
\prime}(0) = m/[1-\xi_1(q_0)]$ (insert to Fig. 1a). When $\xi$ exceeds
$\xi_1(q_0)$, a spherical void of small radius $p_1$ opens in the
center, followed by the first iceberg $p_1 < p < p_2$ (Fig. 1b).
This corresponds to the FGS $\to$ MFS$(1)$ transition, and the values
of $p_1$ and $p_2$ result from the equilibrium and normalization
conditions, Eqs. (4). Their numerical solution shows that the void
radius extends with $\xi$, and in this course the negative effective
mass diminishes and changes to positive. Eventually this again leads
to an instability in the center: $\varepsilon(0) \to \varepsilon (p_1)
= \varepsilon (p_2) = \mu$, when $\xi$ attains the next critical
value $\xi_2(q_0) > \xi_1(q_0)$ (Fig. 1c). But in this case a filled
spherical island (second iceberg) emerges in the center of the void,
and the system passes to MFS$(2)$. With growing $\xi$, the alternating
formation of voids and islands in the center continues infinitely,
and the number of icebergs increases very rapidly, making the
numerical analysis of exact conditions, Eqs. (4), practically
impossible.

However, an effective asymptotical treatment can be proposed for
advanced MFS$(N\gg 1)$, suggested by the general $\xi$-$q_0$ phase
diagram. Considering the monotonically growing function $\xi_1(q_0)$,
Eq. (13), and the fact that $\xi_{N+1}(q_0) > \xi_N(q_0)$, we reasonably
expect that MFS$(N\gg 1)$ are reached with {\it decreasing} screening
parameter $q_0$ at fixed $\xi > 1$. The averaged characteristics of such
state  should be close to the FC characteristics for the same $\xi$.
Below we develop an effective description of MFS$(N\gg 1)$ using this
closeness and prove by the obtained results that such approximation is
asymptotically correct.
\vskip 28 truept

\centerline{\bf 3.  EFFECTIVE DESCRIPTION OF MFS$(N\gg 1)$}
\vskip 12 truept

We expect that, for an MFS$(N\gg 1)$ at given $\xi > 1$ and $q_0 \ll p_f$,
the average occupation function
$$
n_{\rm av}(p) = \int_{p-\Delta}^{p+\Delta}
n(p^\prime){\rm d}p^\prime \eqno(15)
$$
(at $p_f \gg \Delta \gg q_0$), is close to the corresponding FC
function $n_{\rm FC}(p) = \xi^{-1}\theta (p_f - p)$, constant over
the whole range $[0,p_f]$. Therefore we model the ``microscopic''
MFS$(N)$ function $n(p)$ by a certain effective function $n_{\rm
eff}(p)$. In vicinity of any $0 \leq p \leq p_f$ it consists in a
sequence of icebergs of width $\approx\delta(p)/\xi$ with spacers of
width $\approx(1-1/\xi)\delta(p)$ (Fig. 2). The local period
$\delta(p)$ is supposed to be a slow function of $p$ (such that $
\delta^\prime(p)\ll 1$) chosen in order to satisfy the equilibrium
conditions, Eqs. (4), for the effective dispersion law $\varepsilon_
{\rm eff}(p)$. Then the overall multiplicity is asymptotically given by

$$
N=\int_0^{p_f}\frac{{\rm d}p}{\delta(p)}\, . \eqno(16)
$$
We present the effective potential energy as:
$$
\varepsilon_{\rm eff}(p) - \varepsilon_0(p) = V_{\rm eff}(p)=
\sum_{i=1}^N V_i(p)\, , \eqno(17)
$$
where $i$th iceberg with the center in $P_i$ (Fig. 2) contributes by:
$$
V_i(p)=\frac{3\xi}{m} \int_{-\delta_i/2\xi}^{\delta_i/2\xi}(P_i+x)^2
v(p,P_i+x) {\rm d}x\, , \eqno(18)
$$
$\delta_i=\delta(P_i)$. The most important step consists in separating
from $V_{\rm eff}(p)$ the FC function $V_{\rm FC}(p)$, so that
$\varepsilon_{\rm eff}(p) = \mu_{\rm FC} + \delta V(p)$. Then the small
functional $\delta V(p)=V_{\rm eff}(p)-V_{\rm FC}(p)$ can be naturally
approximated by the linear terms in (effectively) small differences
$\delta n(p) = n_{\rm eff}(p) - n_{\rm FC}(p)$ and  $\delta v(p,p^\prime)
= v(p,p^\prime) - v_{\rm FC}(p,p^\prime)$:
$$
\delta V(p)=\delta V_1(p)+\delta V_2(p)\, ,
$$
where
$$
\eqalignno{
\delta V_1(p)=&\frac{3}{m}\int_0^{p_f}{p^\prime}^2
\delta v(p,p^\prime){\rm d}p^\prime \, , \cr
\delta V_2(p)=& \frac{3\xi}{m}\left[\int_0^p{p^\prime}^2\delta n(p^\prime)
{\rm d}p^\prime+\int_p^{p_f}p^\prime\delta n(p^\prime)
{\rm d}p^\prime\right]\, . &(19) \cr}
$$
A direct calculation shows that $\delta V_1(p)$ is a negative
monotonously growing function such that
$$
\delta V_1(0) \approx -\frac{3q_0^2}{2m}\ln \frac{p_f}
{q_0}\, ,\quad \delta V_1(p_f) \approx \frac{\delta
V_1(0)}{2}\, ,
$$
and $V_1(p)\to 0$ at $p \gg p_f$.
When integrating in $\delta V_2(p)$ over each interval $\delta_i$ centered
in $P_i$, we first suppose that the width of $i$th iceberg is exactly
$\delta_i/\xi$. Under this condition, called the geometric equilibrium
(Fig. 2a), the simple rules hold:
$$
\eqalignno{
\int_{-\delta_i/2}^{\delta_i/2}\delta n(P_i+x){\rm d}x&=0\, , \cr
\int_{-\delta_i/2}^{\delta_i/2}\delta n(P_i+x)x&{\rm d}x=0\, , &(20) \cr
\int_{-\delta_i/2}^{\delta_i/2}\delta n(P_i+x)&x^2{\rm d}x=
\delta_i^3\frac{2-\xi^3}{2\xi^3}\, . \cr }
$$
At a special choice $\xi=2^{1/3}$, also the last line in Eq. (20) turns
zero, and the contributions to $V_2(p)$ from all the intervals vanish,
except for that containing $p$ itself. Adopting the above choice, we
obtain the only contribution to $\delta V_2(p)$ from the interval
$P_i-\delta_i \leq p \leq P_i+\delta_i$:
$$
\eqalignno{
V_{\rm osc}(p)=\frac{3\xi}{m}&\int_{-\delta_i/2}^{\delta_i/2}\left
[\frac{P_i+x}{p}\theta(p-P_i-x)+\theta(P_i+x-p)\right]\times \cr
\times&\left[\theta(\frac{\delta_i^2}{2\xi^2}-x^2)-\frac{1}{\xi}\right]
(P_i+x){\rm d}x\,. &(21) \cr }
$$
The totality of such contributions defines a continuous positive function,
reaching zero between the icebergs: $V_{\rm osc}(P_i\pm\delta_i/2)=0$, and
maximum at their centers:
$$
V_{\rm osc}(P_i) = V_{\rm max}(p) \approx \frac{3 \delta^2(p)(\xi-1)}
{4m\xi^2}\, ,
$$
\centerline{\epsfbox{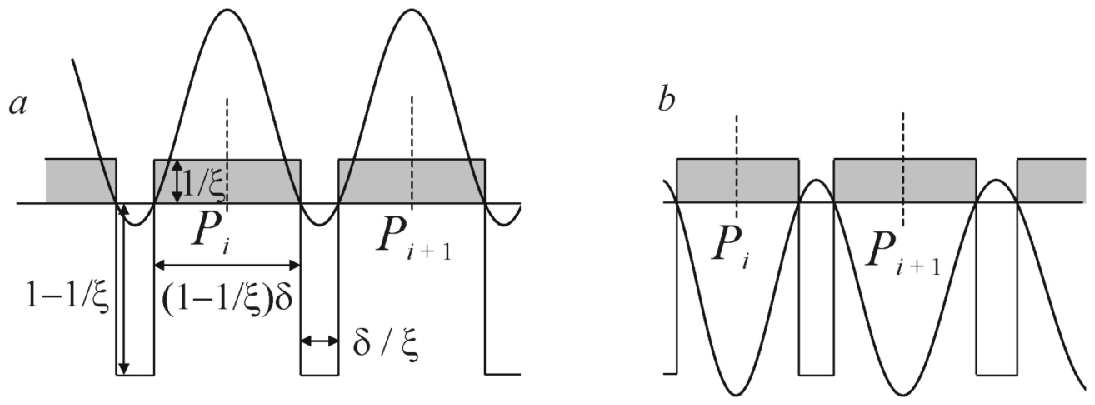}}
{\bf Figure 2.} a) Condition of geometric equilibrium between icebergs
and spacers, b) Physical equilibrium between iceberg distribution and
dispersion law at slow variation of period.
\vskip 12 truept
\noindent
so that its fast oscillating behavior can be modelled by a simple
expression:
$$
V_{\rm osc}(p) =\frac{V_{\rm max}(p)}{2}\left[1+\cos\frac{2\pi p}
{\delta(p)}\right]\, . \eqno(22)
$$
However, the equilibrium for MFS requires that the {\it minima} of
$\varepsilon_{\rm eff}(p)$ be located within icebergs and its {\it
maxima} within spacers, opposite to what happens with $V_{\rm osc}(p)$
at geometric equilibrium (Fig. 2a). This misfit can be corrected with
small correlated shifts of iceberg boundaries from the geometric
equilibrium, described by a finite gradient $\delta^\prime(p)$.
Then non-zero contributions to $\delta V_2(p)$ appear from all the
intervals, giving rise to its slow component $V_{\rm slow}(p) \sim
\delta^\prime(p) p_f^2/m$ while the fast component $V_{\rm osc}(p)$
only obtains a certain phase shift (see below). For this situation,
we present the chemical potential as
$$
\mu=\mu_{\rm FC} + \delta V_1(p) + V_{\rm slow}(p) +  V_b(p) \, ,
  \eqno(23)
$$
where $V_b(p) \sim V_{\rm max}(p)$ is the value of $V_{\rm osc}(p)$ at
iceberg boundaries. The constancy of $\mu$ is mainly controlled by the
condition that $V_{\rm slow}(p)$ scales with $\delta V_1(p)$, leading
to the estimate
$$
\delta^\prime(p) \sim \frac {q_0^2}{p_f^2}
\ln \frac{p_f}{q_0} \, . \eqno(24)
$$
On the other hand, the extrema of $V_{\rm osc}(p)$ estimated, e.g.,
from differentiation of Eq. (22), are shifted due to $\delta^\prime(p)
\neq 0$ by
$$
\sim \frac{\delta^2(p)\delta^\prime(p)}{\delta(p) - p\delta^\prime(p)}\, ,
$$
and they can reach physical equilibrium positions (Fig. 2b) if this
shift is $\sim \delta(p)/2$. This implies the condition on the gradient
$\delta^\prime(p) \approx \delta(p)/p$, agreeing with the initial
assumption $\delta^\prime(p) \ll 1$. Then the comparison with Eq. (24)
gives an estimate for the slow function
$$
\delta(p) \sim p\frac{q_0^2}{p_f^2}\ln \frac{p_f}{q_0}. \eqno(25)
$$
This approximately linear behavior is valid at $p \gg q_0$ and
$p_f - p \gg q_0$, otherwise it is limited by the edge effects, the
limiting values being $\delta_{\rm min} \sim (q_0^3/p_f^2)\ln(p_f/q_0)$
and $\delta_{\rm max} \sim (q_0^2/p_f)\ln(p_f/q_0)$. The $\propto 1/p$
decrease of the density of icebergs, defined by Eq. (25), reflects  the
outward pressure on them from the center. Then the estimate for total
multiplicity follows from Eq. (16):
$$
N \sim \frac{p_f^2}{q_0^2\ln(p_f/q_0)}\int_{q_0}^{p_f}
\frac{{\rm d}p}{p} = \frac{p_f^2}{q_0^2}\, . \eqno(26)
$$
Also, using Eq. (25) in Eq. (22), we confirm that $V_b$ is negligible in
Eq. (23). Correspondingly, the chemical potential $\mu$ is shifted
{\it downwards} with respect to the FC value $\mu_{\rm FC}$ by
$$
\Delta \mu \sim \frac{q_0^2}{m} \ln\frac{p_f}{q_0} \sim
\frac{\mu}{N}\ln N\, , \eqno(27)
$$
and, since the deviations $\sim V_{\rm osc} \ll \Delta\mu$, the energy 
difference between FC and MFS($N$) is orrectly estimated by Eq. (27). 
We notice that all the considerations after Eq. (23) are equally valid 
for the general situation, $\xi \neq 2^{1/3}$.

The asymptotic behavior of MFS($N$) resulting from Eqs. (25-27) depends
only weakly on the coupling constant $\xi$. However this almost universal
regime is only reached for $N$ above some crossover value $N^*$ which
essentially depends on $\xi$. This relates to the fact that multiplicity
$N$ for given $\xi$ grows with $1/q_0$, but this process starts from
$1/q_0 = 1/q_{\xi}$ such that $\xi = \xi_1(q_{\xi})$, Eq. (13). Hence
the universal regime is reached already at $1/q_0 > 1/q_{\xi}$ and
corresponds to $N > N^* \sim (p_f/q_{\xi})^2$. Thus, for $\xi - 1 \ll 1$,
we estimate from Eq. (14) $N^* \sim \ln[1/(\xi - 1)]/(\xi -1)$.
\vskip 28 truept

\centerline{\bf 4.  CONCLUDING REMARKS}
\vskip 12 truept

At $T=0$, our treatment has shown that the energy difference between
FC and MFS($N$) tends to zero when $N$ overcomes the crossover
value $N^*$, $N\geq N^*$. For MFS, the effective mass $M^*$ is defined 
by the dispersion near the boundaries of icebergs:
$$
M^*\sim \frac{p}{\partial\varepsilon_{\rm eff}/\partial p} 
\sim m\frac{p}{\delta(p)}\sim \frac{m N}{\ln N}. \eqno(28)
$$
It is seen from Eq. (28) that at $N\gg N^*$ the effective mass can 
be extremely large, though finite even at $T=0$. Thus MFS represents a 
heavy-fermion system which can be treated as Landau Fermi liquid. 
At finite temperatures, the system persists to be a Landau Fermi 
liquid, but there is a crossover temperature $T^*$ at which the 
difference between FC and MFS vanishes. To calculate $T^*$, we 
observe that the single-particle spectrum should not be altered when 
the temperature reaches $T^*$. The effective mass of a system with FC 
at finite $T$ is given by [15]
$$
M^*\simeq p_{\rm F}{{p_f-p_i}\over{4T}}. \eqno(29)
$$
Upon comparing Eqs. (28) and (29), we obtain
$$
T^*\sim T_{\rm F}\frac{p_f-p_i}{p_{\rm F}}\frac{\ln N}{N}. \eqno(30)
$$
where Fermi temperature $T_{\rm F} = p_{\rm F}^2/2k_{\rm B}m$. At 
$T\geq T^*$, the system comes into the state with the effective mass, 
Eq. (29), and the difference between the MFS and FC states is eliminated. 
Hence, Eq. (30) defines the crossover temperature $T^*$ at which the system 
is caused to pass from a Landau Fermi liquid to a strongly correlated 
Fermi liquid with FC and temperature dependent effective mass. A more 
detailed analysis of the physical properties of this and other model 
Fermi systems and their possible relation to the observed effects in real 
systems will be the topic of forthcoming studies.  
\vskip 28 truept

\centerline{\bf ACKNOWLEDGMENT}
\vskip 12 truept

The present work was partly supported by the Russian Foundation for
Basic Research under Grant No.~01-02-17189.  \vskip 28 truept

\centerline{\bf REFERENCES}
\vskip 12 truept

\item{[1]}
L.~D.~Landau, {Sov.~Phys.~JETP}~{\bf 3}, 920 (1957).
\item{[2]}
I.~Ya.~Pomeranchuk, {Sov.~Phys.~JETP}~{\bf 8}, 361 (1958).
\item{[3]}
J.~Bardeen, L.~Cooper, and J.~Schrieffer, { Phys.~Rev.}~{\bf 108},
1175 (1957).
\item{[4]}
A.~B.~Migdal, { Sov.~Phys.~JETP}~{\bf 5}, 333 (1958).
\item{[5]}
V.~A. Khodel and V.~R.~Shaginyan, { JETP Lett.} {\bf 51}, 553 (1990).
\item{[6]}
P.~Nozieres, { J.~Phys.~France} {\bf 2}, 443 (1992).
\item{[7]}
V.~A. Khodel, V.~R.~Shaginyan, and V.~V.~Khodel, { Phys.~Rep.}~
{\bf 249}, 1 (1994).
\item{[8]}
M. de Llano and J.~P.~Vary, { Phys.~Rev.~C} {\bf 19}, 1083 (1979).
\item{[9]}
M. de Llano, A.~Plastino, and J.~G.~Zabolitsky, { Phys.~Rev.~C}
{\bf 20}, 2418 (1979).
\item{[10]}
S.~A.~Artamonov, Yu.~G.~Pogorelov, and V.~R.~Shaginyan, { JETP Lett.}
{\bf 68}, 897 (1998).
\item{[11]}
M.~V. Zverev and M.~Baldo, { JETP} {\bf 87}, 1129 (1998).
\item{[12]}
M.~V. Zverev, V.~A.~Khodel, and M.~Baldo, { JETP Lett}~
{\bf 72}, 126 (2000).
\item{[13]}
P.~Gegenwart, J.~Clusters, C.~Geibel, K.~Neumaier, T.~Tayama, K.~Tenya,
O.~Trovarelli, and F.~Steglich, { Phys.~Rev.~Lett.} {\bf 89}, 056402 (2002).
\item{[14]}
Yu.G. Pogorelov  and V.R. Shaginyan, cond-mat/0209503.
\item{[15]}
J. Dukelsky, V.A. Khodel, P. Schuck,  and V.R. Shaginyan,
Z. Phys. B {\bf 102}, 245 (1997).

\end